\documentclass[aps,prl,reprint,superscriptaddress]{revtex4-1}
\usepackage{graphicx}
\usepackage{dcolumn}
\usepackage{bm}
\usepackage{amsmath}

\usepackage{amsfonts}
\usepackage{textcomp} 

\usepackage{verbatim}
\usepackage {ulem}

\usepackage{color}
\usepackage{array}
\usepackage{makecell}

\newcolumntype{C}{>{\centering\arraybackslash$}m{2cm}<{$}}

\begin{document}

\title{Lack of correlation between the spin mixing conductance and the ISHE-generated voltages in CoFeB/Pt,Ta bilayers}

\author{A. Conca}

\email{conca@physik.uni-kl.de}

\author{B.~Heinz}

\author{M.~R.~Schweizer}

\author{S.~Keller}

\author{E.~Th.~Papaioannou}

\author{B.~Hillebrands}

\affiliation{Fachbereich Physik and Landesforschungszentrum OPTIMAS, Technische Universit\"at
Kaiserslautern, 67663 Kaiserslautern, Germany}

\date{\today}

\begin{abstract}
We investigate spin pumping phenomena in polycrystalline CoFeB/Pt and CoFeB/Ta bilayers and the correlation between the effective spin mixing conductance $g^{\uparrow\downarrow}_{\rm eff}$ and the obtained voltages generated by the spin-to-charge current conversion via the inverse spin Hall effect in the Pt and Ta layers. For this purpose we measure the in-plane angular dependence of the generated voltages on the external static magnetic field  and we apply a model to separate the spin pumping signal from the one generated by the spin rectification effect in the magnetic layer. Our results reveal a dominating role of anomalous Hall effect for the spin rectification effect with CoFeB and a lack of correlation between $g^{\uparrow\downarrow}_{\rm eff}$ and inverse spin Hall voltages pointing to a strong role of the magnetic proximity effect in Pt in understanding the observed increased  damping. This is additionally reflected on the presence of  a linear dependency of the Gilbert damping parameter on the Pt thickness.

\end{abstract}

\maketitle

\section{Introduction}

In spin pumping experiments,\cite{tser,tser2} the magnetization of a ferromagnetic layer (FM) in contact with a non-magnetic one (NM) is  excited by a microwave field. A  spin current is generated and injected into the NM layer and its magnitude is maximized when the ferromagnetic resonance (FMR) condition is fulfilled. The spin current can be detected by using the inverse spin Hall effect (ISHE) for conversion into a charge current in appropriate materials. The injected spin current $J_{\rm s}$ in the NM layer has the form\cite{tser}

\begin{equation} \label{is}
J_{\rm s}=\frac{\hbar}{4 \pi}g^{\uparrow\downarrow}\hat{m}\times\frac{d\hat{m}}{dt}
\end{equation}
where $\hat{m}$ is the magnetization unit vector and $g^{\uparrow\downarrow}$ is the real part of the spin mixing conductance which  is  controlling the intensity of the generated spin current. Its value is sensitive to the interface properties. The generation of the spin current opens an additional loss channel for the magnetic system and consequently causes an increase in the measured Gilbert damping parameter $\alpha$:

\begin{equation} \label{deltasp}
\Delta\alpha_{\rm sp}= \frac{\gamma\hbar}{4\pi M_s \:d_{\rm FM}}g^{\uparrow\downarrow}
\end{equation}

This expression is  only valid for thick enough NM layers where no reflection of the spin current takes place at the interfaces. In principle, it allows the estimation of $g^{\uparrow\downarrow}$ by measuring the increase in damping compared to the intrinsic value. However, other phenomena, like the magnetic proximity effect (MPE) in the case of Pt or interface effects depending on the exact material combination or capping layer material, can have the same influence, \cite{fept,ana} which challenges the measurement of the contribution from the spin pumping. In this sense, it is preferable to use an effective value $g^{\uparrow\downarrow}_{\rm eff}$. Still, if the spin pumping is the main contribution to the increase in $\alpha$, a correlation between $g^{\uparrow\downarrow}_{\rm eff}$ and the measured ISHE voltages is expected. A suitable approach in order to understand  the weight of MPE on the value of $g^{\uparrow\downarrow}_{\rm eff}$ is the use of FM/NM with varying NM metals, with presence and absence of the MPE effect. The measurement of $\Delta\alpha$ and $g^{\uparrow\downarrow}_{\rm eff}$ together with the ISHE voltages generated by the spin current in the NM layer can bring clarity to the issue. 

However, the generation of an additional dc voltage by the spin rectification effect,\cite{harder,soh,gui,zhang} which adds to the voltage generated by the ISHE spin-to-charge conversion,  deters the analysis of the obtained data. The spin rectification  originates from the precession of the magnetization in conducting layers with magnetoresistive properties, mainly Anisotropic Magnetoresistance (AMR) and Anomalous Hall Effect (AHE). Information about the physics behind the measured voltage can only be obtained after separation of the different contributions. For this purpose, we made use of the different angular dependencies of the contributions under in-plane rotation of the external magnetic field. 

\begin{figure}[b]
    \includegraphics[width=1.0\columnwidth]{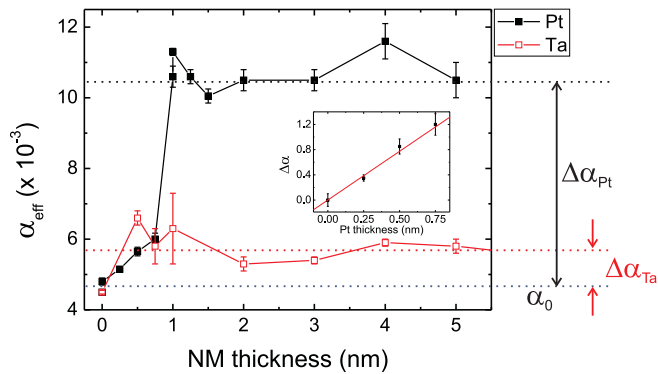}
	  \caption{\label{dampingptta}(Color online) Dependence of the effective Gilbert damping parameter $\alpha_{\rm eff}$ on the thickness of the NM metal. A large increase in damping is observed for the Pt case while a very small but not vanishing increase is observed for Ta. From the change $\Delta \alpha$ the effective spin mixing conductance $g^{\uparrow\downarrow}_{\rm eff}$ is estimated using Eq.~\ref{deltasp}.}
\end{figure}

\section{Experimental details}

Here, we report on results on polycrystalline Co$_{40}$Fe$_{40}$B$_{20}$/Pt,Ta bilayers grown by rf-sputtering on Si substrates passivated with SiO$_2$. CoFeB is a material choice for the FM layer due to its low damping properties and easy deposition.\cite{cofeb-alpha,cofeb-ann} A microstrip-based VNA-FMR setup was used to study the damping properties. A more detailed description of the FMR measurement and analysis procedure is shown in previous work.\cite{fept,cofeb-ann} A quadrupole-based lock-in setup described elsewhere\cite{keller}  was used in order to measure the ISHE generated voltage. The dependence of the voltage generated during the spin pumping experiment on the in-plane static external field orientation is recorded for a later separation of the pure ISHE signal from the spin rectification effect.

\section{Gilbert damping parameter and spin mixing conductance}

Figure~\ref{dampingptta} shows the dependence of the effective damping parameter $\alpha_{\rm eff}$ (sum of all contributions) on the thickness $d$ of the NM metal for a CoFeB layer with a fixed  thickness of 11~nm. The case $d=0$~nm represents the case of reference layers with Al capping. From  previous studies it is known that the use of an Al capping layer induces a large increase of damping in Fe epitaxial layers.\cite{fept} For polycrystalline NiFe and CoFeB layers this is not the case and it allows the measurement of the intrinsic value $\alpha_{\rm 0}$. \cite{ana}

The observed behavior differs strongly for Pt and Ta. In the Pt case a large increase in damping is observed with a sharp change around  $d=1$~nm and a fast saturation for larger thicknesses. This is qualitatively very similar to our previous report on Fe/Pt bilayers.\cite{fept}  From the measured $\Delta\alpha$ we extract the value $g^{\uparrow\downarrow}_{\rm eff}=6.1 \pm 0.5 \cdot 10^{19}$m$^{-2}$. This value is larger than the one reported previously in our group\cite{ana} for thinner CoFeB layers with larger intrinsic damping $4.0 \pm 1.0 \cdot 10^{19}$m$^{-2}$  and also larger than the value  reported  by Kim \textit{et al.} \cite{kim}, $5.1\cdot 10^{19}$m$^{-2}$. The impact of the Ta layer on damping is very reduced and, consequently, a low value for $g^{\uparrow\downarrow}_{\rm eff}$ of $0.9 \pm 0.3 \cdot 10^{19}$m$^{-2}$ is obtained. This value is now smaller than the one reported by Kim \textit{et al.} $1.5 \cdot 10^{19}$m$^{-2}$) indicating that the difference between CoFeB/Pt and Ta is larger in our case. A reference has also to be made to the work of Liu \textit{et al.} on CoFeB films thinner than in this work. \cite{liu} There, no value for the spin mixing conductance is provided, but the authors claim a vanishing impact on $\alpha$ for the Ta case. On the contrary the increase due to Pt is almost three times  larger than ours, pointing to a huge difference between both systems. In any case, the trend is similar, only the relative difference between Ta and Pt changes. 

\begin{figure}[b]
    \includegraphics[width=0.7\columnwidth]{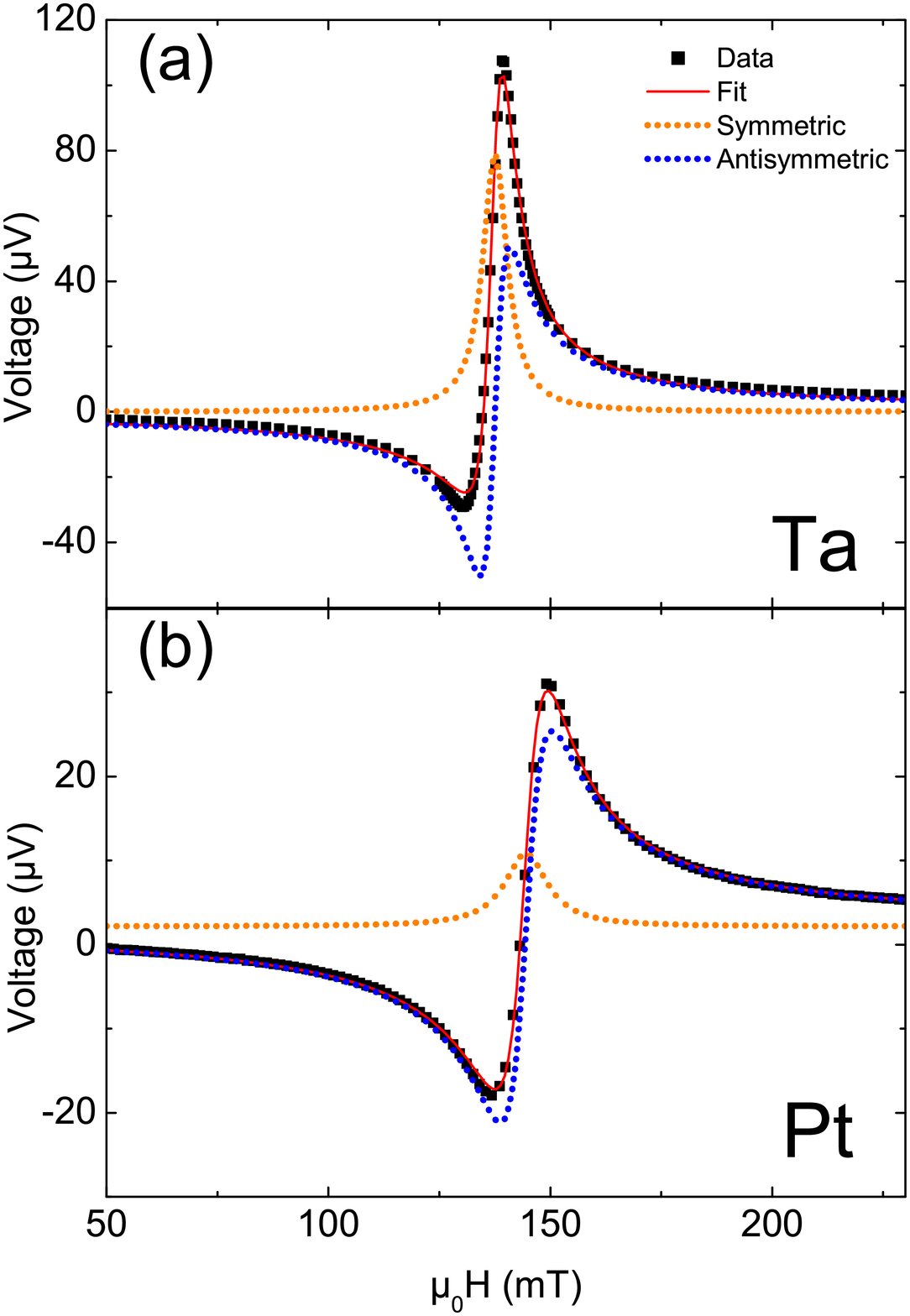}
	  \caption{\label{spectra}(Color online) Voltage spectra measured for (a) CoFeB/Ta and (b) CoFeB/Pt at 13~GHz. The solid line is a fit to Eq.~\ref{fitspectra}. The symmetric voltage $V_{\rm sym}$ and antisymmetric voltage $V_{\rm antisym}$ contributions are separated and plotted independently (dashed lines). The voltage signal is dominated by  $V_{\rm antisym}$ in the Pt case and by $V_{\rm sym}$ in the Ta case. }
\end{figure}

A closer look to the data allows to distinguish a  region in the Pt damping evolution  prior to the sharp increase  where  a linear behavior is recognized ($d<1$~nm). A linear thickness  dependence of $\alpha$ in spin-sink ferromagnetic films and in polarized Pt has been reported. \cite{caminale,gosh} The increase in damping due to spin current absorption in the Pt with ferromagnetic order can then be described by: 

\begin{equation} \label{mpe}
\Delta\alpha= \Delta\alpha_{\rm MPE}\cdot d_{\rm Pt}/d^{\rm Pt}_{\rm c}
\end{equation}
where $\Delta\alpha_{\rm MPE}$ is the total increase in damping due only to the magnetic proximity effect in Pt, $d_{\rm Pt}$ is the thickness of the Pt layer and $d^{\rm Pt}_{\rm c}$ is a cutoff thickness which is in the order of magnitude of the coherence length in ferromagnetic layers.\cite{gosh,stiles}

The inset in Fig.~\ref{dampingptta} shows a fit of Eq.~\ref{mpe} from where   $d^{\rm Pt}_{\rm c}=0.8$~nm is obtained assuming a value $\Delta\alpha_{\rm MPE} =1.2$. The  value is in qualitative agreement with the  reported thickness where MPE is present in Pt, ($d^{\rm Pt}_{\rm MPE}\leq 1$~nm \cite{suzuki,wilhelm}) and is lower than the one reported for Py/Pt systems.\cite{caminale} 

The increase of damping due to spin pumping is described by an exponential dependence and explains the sharp increase at $d_{\rm Pt}=1$~nm. However, the fast increase does not allow for a deep analysis and it is pointing to a spin diffusion length in Pt not larger than 1~nm.

In any case, this point has to been treated with care. The contribution of MPE to damping can be easily underestimated and consequently also the value for $d^{\rm Pt}_{\rm c}$. In any case, the value can be interpreted as a lower limit for $\Delta\alpha_{\rm MPE}$. If this is  substracted, under the  assumption that the rest of increase is due to spin pumping, the spin mixing conductance due only to the this effect would be $g^{\uparrow\downarrow}_{\rm eff}=4.9 \pm 0.5 \cdot 10^{19}$m$^{-2}$.

\section{Electrical detection of  spin pumping }

\begin{figure}[b]
    \includegraphics[width=0.6\columnwidth]{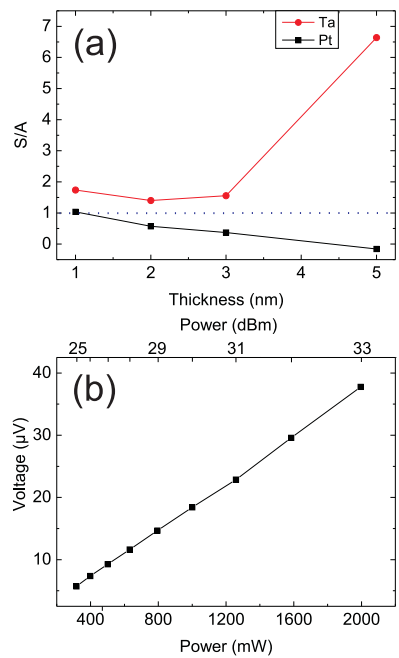}
	  \caption{\label{ratio-linear}(Color online)  (a) Dependence of the ratio S/A $= V_{\rm sym}/V_{\rm antisym}$ on the thickness of the NM layer. (b) Dependence of the total voltage on the applied microwave power proving the  measurements were carried out in the linear regime.}
\end{figure}

Figure~\ref{spectra}(a),(b) shows two voltage measurements recorded at 13~GHz for a NM thickness of 3~nm and a nominal microwave power of 33~dBm. The measured voltage is the sum of the contribution of the ISHE effect and of spin rectification effect originating from the different magnetoresistive phenomena in the ferromagnetic layer. While the spin rectification effect generates both a symmetric and an antisymmetric contribution, \cite{harder,soh,gui} the pure ISHE signal is only symmetric. For this reason a separation of both is carried out by fitting the voltage spectra (solid line) to

\begin{equation} \label{fitspectra}
\begin{split}
V_{\rm meas} &= V_{\rm sym}\frac{(\Delta H)^2}{(H-H_{\rm FMR})^2+(\Delta H)^2}+ \\
&+ V_{\rm antisym}\frac{-2\Delta H (H-H_{\rm FMR})}{(H-H_{\rm FMR})^2+(\Delta H)^2}
\end{split}
\end{equation}
where $\Delta H$ and $H_{\rm FMR}$ are the linewidth and the resonance field, respectively. The dotted lines in Fig.~\ref{spectra} show the two contributions. When comparing the data for Pt and Ta some differences are observed. First of all, the absolute voltage values are smaller for the Pt cases and, more important, the relative weight of both contributions is different. While the first point is related to the different conductivity of Ta and Pt, the second one is related to the intrinsic effect causing the voltage. We calculate the ratio S/A $= V_{\rm sym}/V_{\rm antisym}$ for all the measurements and the results are shown in Fig.~\ref{ratio-linear}(a) as a function of the NM thickness. While the antisymmetric contribution is dominating in the Pt samples with a S/A ratio  smaller than 1 for the samples with Pt, the opposite is true for the Ta case.
 Since the ISHE signal is contributing only to $V_{\rm sym}$ it might be concluded that spin pumping is taking place stronger in the Ta system. However, since also the spin rectification effect has a symmetric contribution, this conclusion cannot be supported. Furthermore, since the spin Hall angle $\theta _{\rm SHE}$ has opposite sign in these two materials, also the ISHE signal should have it. In apparent contradiction to this, we observe that both symmetric contributions have the same sign in (a) and (b). This  points to the fact that for Pt, $V_{\rm sym}$ is dominated by the spin rectification effect, which does not change sign and overcompensates a smaller ISHE signal. All these considerations have the consequence that it is not possible to extract complete information of the origin of the measured voltage by analyzing single spectra. For the same reason, the large increase in S/A for Ta for $d=5$~nm or the change in sign for Pt with the same thickness cannot be correctly explained until the pure ISHE signal is not separated from the spin rectification effect. As already pointed out in recent papers\cite{harder,soh,gui,saitoh,keller}, an analysis of the angular dependence (in-plane or out-of-plane) of the measured voltages can be used to separate the different contributions.

In any case, before proceeding it has to be proven that all the measurements were performed in the linear regime with small cone angles for the magnetization precession. The measurements performed out of this regime would have a large impact on the linewidth and a Gilbert-like damping would not be guaranteed. Figure~\ref{ratio-linear}(b) shows the dependence of the voltage amplitude on the microwave nominal power proving indeed that the measurements were carried on in the linear regime.

\section{Separation of the ISHE signal from the spin rectification voltage}

We performed in-plane angular dependent measurements of the voltage and Eq.~\ref{fitspectra} was used to extract $V_{\rm sym,antisym}$ for each value of the azimuthal angle $\phi$ spanned between the direction of the magnetic field and the microstrip antenna used to excite the magnetization. We used a model based on the work of Harder \textit{et al.}\cite{harder} to fit the dependence. This model considers two sources for the spin rectification, which are the Anisotropic Magnetoresistance (AMR) and the Anomalous Hall Effect (AHE):

\begin{equation} \label{model}
\begin{split}
V_{\rm sym} &= V_{\rm sp}\,\text{cos}^3(\phi)+\\
 &+ V_{\rm AHE}\,\text{cos}(\Phi)\text{cos}(\phi)+V^{\rm sym}_{\rm AMR-\perp}\,\text{cos}(2\phi)\text{cos}(\phi)\\
 &+V^{\rm sym}_{\rm AMR-\parallel}\,\text{sin}(2\phi)\text{cos}(\phi) \\
V_{\rm antisym} &=  V_{\rm AHE}\,\text{sin}(\Phi)\text{cos}(\phi)\,\,+V^{\rm antisym}_{\rm AMR-\perp}\,\text{cos}(2\phi)\text{cos}(\phi)\\
&+V^{\rm antisym}_{\rm AMR-\parallel}\,\text{sin}(2\phi)\text{cos}(\phi)  
\end{split}
\end{equation}

Here, $V_{\rm sp}$ and $V_{\rm AHE}$ are the contributions from spin pumping (pure ISHE) and from AHE, respectively. $\Phi$ is the phase  between the rf electric and magnetic fields in the medium. The contribution from the AMR is divided in one generating a transverse $\perp$ (with respect to the antenna) or longitudinal $\parallel$ voltage. In an ideal case with perfect geometry and point-like electrical contacts $V^{\rm sym,antisym}_{\rm AMR-\parallel}$ should be close to zero.

\begin{figure}[t]
    \includegraphics[width=0.7\columnwidth]{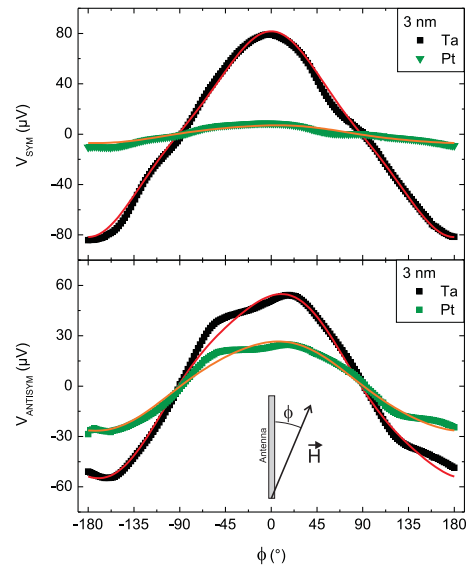}
	  \caption{\label{angular}(Color online) Angular dependence of $V_{\rm sym}$ (top) and $V_{\rm antisym}$ (bottom) for CoFeB/Pt,Ta  samples with NM thickness of 3~nm. The lines are a fit to the model described in Eq.~\ref{model}.}
\end{figure}

\begin{figure}[t]
    \includegraphics[width=0.7\columnwidth]{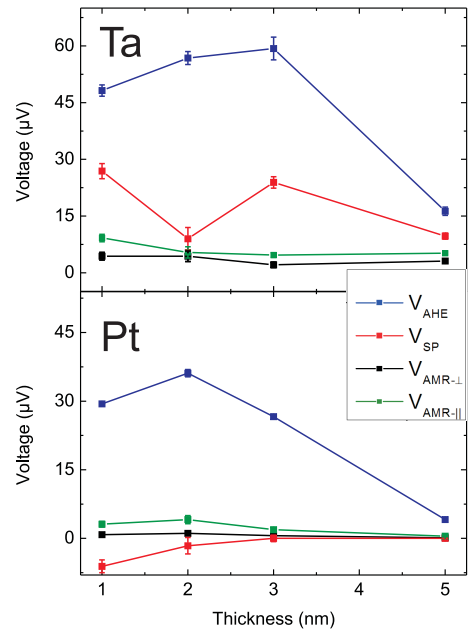}
	  \caption{\label{thickness}(Color online) NM thickness dependence of the different contributions to the measured voltages extracted from the angular dependence of $V_{\rm sym}$ and $V_{\rm antisym}$ for Ta (top) and Pt (bottom).}
\end{figure}

Figure~\ref{angular} shows the angular dependence of $V_{\rm sym}$ (top) and $V_{\rm antisym}$ (bottom) for the  samples with NM thickness of 3~nm. The lines are a fit to the model which is able to describe the dependence properly. From the data it can be clearly concluded that while the values of $V_{\rm antisym}$ are comparable, with the difference resulting from the different resistivity of Pt and Ta,  the values of $V_{\rm sym}$ are  much larger for Ta. The values obtained from the fits for the different contributions are plotted in Fig.~\ref{thickness} as a function of the thickness of the NM layer. The value of $\Phi$ is ruling the lineshape of the electrically measured FMR peak\cite{harder2} which is always a combination of a dispersive ($D$, antisymmetric) and a Lorentzian ($L$, symmetric) contribution in the form $D+iL$.  In order to compare the relative magnitudes of the different contributions independently of $\Phi$ we compute the quantities $V_{\rm AMR-\parallel,\perp}=\sqrt{\left(V^{\rm antisym}_{\rm AMR-\parallel,\perp}\right)^2+\left(V^{\rm sym}_{\rm AMR-\parallel,\perp}\right)^2}$  which it is equivalent to $\sqrt{D^2+L^2}$ and we show them together with $V_{\rm AHE}$ and $V_{\rm sp}$. This step is important to allow for comparison of the different contributions independent of the value of $\Phi$.

Several conclusions can be extracted from Fig.~\ref{thickness}. First of all, the spin rectification effect in CoFeB systems is almost fully dominated by the AHE. AMR plays a very minor role. This is a difference with respect to NiFe or Fe. \cite{soh,harder2,keller} This is correlated with the very large AHE reported in CoFeB films. \cite{zhu,zhu2} Second, the voltages generated by the spin pumping via the ISHE are larger in the case of Ta and of opposite sign as expected from the different sign of $\theta _{\rm SHE}$ in both materials. This solves the apparent contradiction observed by the positive symmetric contributions in both materials as shown in Fig.~\ref{spectra}(a) and (b) and confirms the interpretation than in the case of Pt the symmetric contribution is dominated by the spin rectification effect with opposite sign to the ISHE signal. Again, this shows that the interpretation using single spectra may lead to confusion and that angle dependent measurements are required.

The evolution of the spin rectification voltages with NM thickness shows a saturation behavior in both cases for small thicknesses and a decrease with the NM layer thickness compatible with a dominant role of the resistance of the CoFeB layer. This is expected from the resistivity values for amorphous CoFeB layers, 300-600~$\mu$m$\cdot $cm,\cite{jen} which are much larger than for $\beta$-Ta (6-10~$\mu$m$\cdot $cm) or sputtered Pt (100-200~$\mu$m$\cdot $cm).\cite{stella,sagasta} However, the dependence does not completely agree with the expected behavior\cite{saitoh} $1/d_{\rm NM}$ pointing out to additional effects like a variation of the conductivity of Pt for the  thinner layers.

Concerning the correlation  of the absolute values of the ISHE-generated voltages and  the spin Hall angles in both materials, unfortunately the scatter in $\theta _{\rm SHE}$ values in the literature is very large.\cite{sinova} However this is reduced if we consider works were $\theta _{\rm SHE}$ was measured simultaneously for Pt and Ta in similar samples. In YIG/Pt,Ta systems\cite{wang,hahn} it was determined that $\lvert \theta^{\rm Pt} _{\rm SHE} \rvert > \lvert \theta^{\rm Ta} _{\rm SHE} \rvert $ with a relative difference of around 30$\%$ which it is at odds with our results. On the contrary, in CoFeB/Pt,Ta bilayers $\lvert \theta^{\rm Ta} _{\rm SHE} \rvert=0.15 > \lvert \theta^{\rm Pt} _{\rm SHE} \rvert=0.07 $ is reported.\cite{liu} However the difference is not large enough to cover completely the difference in our samples. In order to explain this point together with  the absolute low value in CoFeB/Pt we have to take into account the possibility of a certain loss of spin current at the interface FM/Pt or at the very first nanometer, the latter due to the presence of a static magnetic polarization due to the proximity effect. With this the spin current effectively being injected in Pt would be lower than in the Ta case.

The data does not allow for a quantitative estimation of the spin diffusion length $\lambda_{\rm sd}$, but in any case the evolution is only compatible with a value for Pt not thicker than 1~nm, similar to reported values for sputtered Pt\cite{sagasta} and a a value of a few nm for Ta, also compatible with literature.\cite{hahn}

 An important point is the lack of correlation of $g^{\uparrow\downarrow}_{\rm eff}$ and the expected generated spin current using Eq.~\ref{is} with the absolute measured ISHE voltage that results from the spin-to-charge current conversion, obtained after the separation from the overimposed spin rectification signal.   This is true even if we substract the MPE contribution assumed for Eq.~\ref{mpe}. The same  non-mutually excluding explanations are  possible here:  $\Delta\alpha$ in Pt in mainly due to the MPE, or the spin current pumped into Pt vanishes at or close to the interface. The first alternative would render Eq.~\ref{deltasp} unuseful since most of the increase in damping is not due to spin pumping as long as the MPE is present. The second would reduce the validity of Eq.~\ref{is} to estimate the current injected in Pt and converted into a charge current by the ISHE.  In any case, CoFeB/Ta shows very interesting properties, with strong spin pumping accompanied by only a minor impact on $\alpha$.
 

Let us discuss the limitations of the model defined in Eq.~\ref{model} and the suitability to describe the measurements. First of all, the model assumes a perfect isotropic material. The anisotropy in CoFeB is known to be small but not zero and a weak uniaxial anisotropy is present. The effect on the angular dependence is negligible. The model assumes also a perfect geometry and point-like electrical contacts to measure the voltages. Our contacts are extended ($\sim$200~$\mu$m) and a small misalignment is possible (angle between the antenna and the imaginary line connecting the electrical contacts may not be exactly 90$^{\circ}$). This is the most probable reason for the non-vanishing small value for $V^{\rm sym,antisym}_{\rm AMR-\parallel}$. Nevertheless, the angular dependence of the measured voltage is well described by the model and no large deviations are observed. 

\section{Conclusions}

In summary, we made use of in-plane angular dependent measurements to separate ISHE-generated from spin rectification voltages and we compare the absolute values and thickness dependence for Pt and Ta. Differently to other materials, the spin rectification signal in CoFeB is almost fully dominated by AHE. No correlation between the observed spin mixing conductance  via FMR measurement and the spin pumping signal is obtained pointing to a dominant role of the magnetic proximity effect in the increase in damping  with Pt.

\section*{Acknowledgements}

Financial support by  M-era.Net through the HEUMEM project and by the Carl Zeiss Stiftung  is gratefully acknowledged.

\end{document}